\documentclass
[superscriptaddress,secnumarabic,amssymb,amsmath,nobibnotes,aps,prd,showkeys,showpacs,nofootinbib,nopacsnumber,onecolumn,12pt]{revtex4}%
\usepackage{graphicx}
\usepackage{epsf}
\usepackage{bm}
\usepackage{amsmath}
\usepackage{amsfonts}
\usepackage{amssymb}%
\providecommand{\U}[1]{\protect\rule{.1in}{.1in}}

\newcommand{\be}{\begin{equation}}
\newcommand{\ee}{\end{equation}}

\newcommand{\mincir}{\raise
-3.truept\hbox{\rlap{\hbox{$\sim$}}\raise4.truept\hbox{$<$}\ }}
\newcommand{\magcir}{\raise
-3.truept\hbox{\rlap{\hbox{$\sim$}}\raise4.truept\hbox{$>$}\ }}

\begin{document}
\title{Anisotropic spacetimes in $f(T,B)$ theory IV: Noether symmetry analysis}
\author{Andronikos Paliathanasis}
\email{anpaliat@phys.uoa.gr}
\affiliation{Institute of Systems Science, Durban University of Technology, Durban 4000,
South Africa }
\affiliation{Instituto de Ciencias F\'{\i}sicas y Matem\'{a}ticas, Universidad Austral de
Chile, Valdivia 5090000, Chile}

\begin{abstract}
The Noether symmetry analysis is applied for the analysis of the field
equations in an anisotropic background in $f(T,B)$-theory. We consider the
$f\left(  T,B\right)  =T+F\left(  B\right)  $ which describes a small
deviation from TEGR introduced by the boundary scalar $B$. For the Bianchi\ I,
Bianchi III and Kantowski-Sachs geometries there exists a minisuperspace
description and Noether's theorems are applied. We investigate the existence
of invariant point transformations. We find that for the Bianchi I spacetime
the gravitational field equations are Liouville integrable for the $F\left(
B\right)  =-\frac{B}{\lambda}\ln B$ theory. The analytic solution is derived
and the application of Noether symmetries to the Wheeler-DeWitt equation of
quantum cosmology is discussed.

\end{abstract}
\date{\today}
\maketitle

\section{Introduction}

\label{sec1}

The need to explain cosmological observations \cite{Teg,Kowal,Komatsu,Ade15}
has led the scientific society the introduction of new dynamical degrees of
freedom in Einstein's field equations \cite{clifton}. The main idea is the new
degrees of freedom to drive the dynamics of the cosmological evolution such
that the observable phenomena can be explained. These additional degrees of
freedom correspond to exotic matter source components or they describe
geometrodynamical degrees of freedom which are the result of the modification
of the gravitational Action Integral
\cite{Ratra,Sotiriou,odin1,star,fere,Overduin}.

General Relativity and its modifications are non-linear theories. Thus, in
order to extract important physical information from a specific gravitational
theory methods from the study of dynamical systems in Analytic Mechanics
should be applied. Some powerful methods of Analytic Mechanics which have been
applied to gravitational physics are the analysis of the stationary points
\cite{DynSystemsWain,DynSystemsColey,dyn1,dyn2,dyn3}, the Eisehnhart lift
\cite{ein1,ein2}, the singularity analysis \cite{sin1,sin2} and the symmetry
analysis \cite{sym1,sym3}. The two latter methods are related to the
determination of exact and analytic solutions and to the construction of
invariant manifolds in the space of the dynamical variables.

The study for the existence and the determination of exact and analytic
solutions for a given dynamical system is an essential approach to the
investigation of any dynamical system in applied mathematics. Exact and
analytic solutions when they exist for a dynamical system indicate that there
are actual solutions which describe the trajectories of the dynamical system.
We can study in details the behaviour of the trajectories, i.e. solutions, and
understand the initial value problem. Dynamical systems which admit exact and
analytic solutions can be used as important toy models for the study of the
real world phenomena. For instance, the well-known oscillator is a simple
integrable dynamical system which has applications in all areas of physics.

A powerful approach for the study of cosmological theories is the Noether
symmetry analysis \cite{ns1,ns2,ns3,ns4,ns5,ns6,ns7,ns8,ns9,ns10}. In 1918
\cite{em1}, Emmy Noether published two theorems which connect transformations
of the Action Integral with the invariance of the variational principle and
the existence of invariant functions known as conservation laws. Noether's
work introduces a simple and systematic way for the construction of
conservation laws for non-linear differential equations. It has been widely
applied for the study of various cosmological models and inspired by the work
of Ovsiannikov \cite{ovs}, Noether symmetry analysis has been used to classify
the unknown functions and parameters of a given gravitational model according
to the admitted Noether algebra. However, from the discussion in \cite{bas1}
we can conclude that the Noether classification scheme has a geometric
character related to that of gravity. The symmetry vectors for the field
equations are generated by the minisuperspace, the geometric space, where the
dynamical variables are defined.

This piece of work is the last of a series of studies on the analysis of
anisotropic spacetimes in modified teleparallel $f\left(  T,B\right)  $-theory
\cite{bh1,revtel}. In previous studies, we investigated the dynamical
behaviour of the gravitational field equations in Bianchi I \cite{paper1}, in
Kantowski-Sachs \cite{paper2} and in Bianchi III geometries \cite{paper3}. We
focused on the $f\left(  T,B\right)  =T+F\left(  B\right)  $ theory and we
investigate if the specific theory can solve the flatness and isotropic
problems when the initial conditions are that of anisotropic spacetimes. We
found that for all the possible sets of initial conditions there always exists
an attractor which describes a spatially flat accelerated FLRW Universe.

$f\left(  T,B\right)  $-theory is a fourth-order theory and with the use of a
Lagrange multiplier the higher-order derivatives can be attributed to a scalar
field such that the field equations are of second-order but with additional
dynamical variables \cite{an1}. In addition for the anisotropic spacetimes of
our consideration, the field equations have a minisuperspace description which
means that the Noether symmetry analysis can be applied. There are various
studies in the literature on the application of Noether symmetries in
teleparallel theory and its modifications, see for instance
\cite{nt1,nt2,nt3,nt4,nt6}. In \cite{nt4} the Noether symmetries investigated
in $f\left(  T,B\right)  $-theory for the Bianchi\ I spacetime, however only a
special case of Noether's theorem was applied and the Noether analysis is not
complete. The plan of the paper is as follows.

In Section \ref{sec2} we present the basic properties and definitions of
Noether's work. In Section \ref{sec3}, we present the minisuperspace
description for the field equation in $f\left(  T,B\right)  $-theory for a
background geometry described by the Bianchi I, the Kantowski-Sachs or the LRS
Bianchi III spacetime. The Noether symmetry analysis is presented in Section
\ref{sec4} where we show how the symmetry vectors can be applied in order to
derive exact solutions. Finally,our conclusions are given in Section
\ref{sec5}.

\section{Noether symmetries}

\label{sec2}

In this Section we present the basic results of Noether's work for
one-parameter point transformations. Because in this work we deal with
second-order differential equations we focus on Lagrangian functions of the
form $L\left(  t,\mathbf{q},\mathbf{\dot{q}}\right)  $,~where $t$ is the
independent variable and $\mathbf{q}$ remarks the dependent variables, where
$\mathbf{\dot{q}}=\frac{d\mathbf{q}}{dt}$.

Assume now the Action Integral \cite{em1}%
\begin{equation}
A=\int_{t_{0}}^{t_{1}}L\left(  t,\mathbf{q,\dot{q}}\right)  dt,
\end{equation}
and the one-parameter infinitesimal transformation%
\begin{equation}
\bar{t}=t+\varepsilon\xi\left(  t,\mathbf{q}\right)  ,~\mathbf{\bar{q}%
}=\mathbf{q}+\varepsilon\mathbf{\eta}\left(  t,\mathbf{q}\right)  \label{4422}%
\end{equation}
where $\varepsilon$ is an infinitesimal parameter, i.e. $\varepsilon
^{2}\rightarrow0$. \ The generator vector field for the latter infinitesimal
transformation is $X=\tau\partial_{t}+\eta\partial_{\mathbf{q}}$

Under the application of the transformation (\ref{4422}) the Action integral
reads%
\begin{equation}
\bar{A}=\int_{\bar{t}_{0}}^{\bar{t}_{1}}L\left(  \bar{t},\mathbf{\bar{q}%
,\dot{\bar{q}}}\right)  d\bar{t}%
\end{equation}
where now $\mathbf{\dot{\bar{q}}}$ is defined as $\mathbf{\dot{\bar{q}}}%
=\frac{d\mathbf{\bar{q}}}{d\bar{t}}$.

By expanding around $\varepsilon^{2}\rightarrow0$, the Action Integral
$\bar{A}$ becomes%
\begin{equation}
\bar{A}=A+\varepsilon\int_{t_{0}}^{t_{1}}\left(  \xi{\frac{\partial
L}{\partial t}}+\mathbf{\eta}{\frac{\partial L}{\partial\mathbf{q}}%
}+\mathbf{\zeta}{\frac{\partial L}{\partial\mathbf{\dot{q}}}}+\dot{\xi
}L\right)  dt+\varepsilon F,
\end{equation}
where $F=-\int_{t_{0}}^{t_{1}}\dot{f}dt$ and $\mathbf{\zeta}=\mathbf{\dot
{\eta}}-\mathbf{\dot{q}}\dot{\xi}$

Thus, the Action Integral remains invariant under the application of the
infinitesimal transformation (\ref{4422}) if and only if $\bar{A}=A$, that is
\cite{em1}%
\begin{equation}
\xi{\frac{\partial L}{\partial t}}+\mathbf{\eta}{\frac{\partial L}%
{\partial\mathbf{q}}}+\zeta{\frac{\partial L}{\partial\mathbf{\dot{q}}}}%
+\dot{\xi}L=\dot{f}\label{sm1}%
\end{equation}
The latter expression is Noether's first theorem and the generator $X$ is
called Noether symmetry for the Lagrangian function $L\left(  t,\mathbf{q}%
,\mathbf{\dot{q}}\right)  $.

Noether's second theorem indicates that for every Noether symmetry $X$ there
exists a function $I\left(  t,\mathbf{q,\dot{q}}\right)  $ defined as
\cite{em1}
\begin{equation}
I\left(  t,\mathbf{q,\dot{q}}\right)  =f-\left[  \xi L+\left(  \mathbf{\eta
}-\mathbf{\dot{q}}\xi\right)  \frac{\partial L}{\partial\mathbf{\dot{q}}%
}\right]  \label{sm2}%
\end{equation}
such that $I\left(  t,\mathbf{q,\dot{q}}\right)  $ is a conservation law, that
is $\dot{I}\left(  t,\mathbf{q,\dot{q}}\right)  =0$. For more details on
Noether's theorem and extensions we refer the reader to the recent discussion
in \cite{nsd1}.\hfill

\section{Anisotropic spacetimes in $f\left(  T,B\right)  $-theory}

\label{sec3}

In $f\left(  T,B\right)  $-theory the fundamental geometric invariants is the
$T$ is the torsion scalar for the Weitzenb{\"{o}}ck connection and the
boundary term$~B=2e^{-1}\partial_{\nu}\left(  eT_{\rho}^{~\rho\nu}\right)  $.
The gravitational Action Integral is \cite{bh1}
\begin{equation}
S_{f\left(  T,B\right)  }=\frac{1}{16\pi G}\int d^{4}xef\left(  T,B\right)
\label{cc.06}%
\end{equation}
in which the field equations are%

\begin{align}
0  &  =ef_{,T}G_{a}^{\lambda}+\left[  \frac{1}{4}\left(  Tf_{,T}-f\right)
eh_{a}^{\lambda}+e(f_{,T})_{,\mu}S_{a}{}^{\mu\lambda}\right] \nonumber\\
&  +\left[  e(f_{,B})_{,\mu}S_{a}{}^{\mu\lambda}-\frac{1}{2}e\left(
h_{a}^{\sigma}\left(  f_{,B}\right)  _{;\sigma}^{~~~;\lambda}-h_{a}^{\lambda
}\left(  f_{,B}\right)  ^{;\mu\nu}g_{\mu\nu}\right)  +\frac{1}{4}%
eBh_{a}^{\lambda}f_{,B}\right]  . \label{cc.08}%
\end{align}

The three anisotropic spacetimes of our consideration, Bianchi I,
Kantowski-Sachs and LRS\ Bianchi III are described by the line element
\begin{equation}
ds^{2}=-N^{2}\left(  t\right)  dt^{2}+e^{2\alpha\left(  t\right)  }\left(
e^{2\beta\left(  t\right)  }dx^{2}+e^{-\beta\left(  t\right)  }\left(
dy^{2}+f^{2}\left(  y\right)  dz^{2}\right)  \right)  \label{ch.03}%
\end{equation}
where~$f\left(  y\right)  =1$ is for Bianchi I space,$~f\left(  y\right)
=\sin\left(  y\right)  $ for the Kantowski-Sachs geometry and $f\left(
y\right)  =\sinh\left(  y\right)  $ for the Bianchi III spacetime. Parameters
$\alpha\left(  t\right)  $ and $\beta\left(  t\right)  $ are the two scale
factors, specifically $\alpha\left(  t\right)  $ remarks the radius of the
three-dimensional hypersurface while $\beta\left(  t\right)  $ is the
anisotropic parameter.

In teleparallelism we should define the proper frame for each spacetime in
order the limit of General Relativity to be recovered. Indeed, following
\cite{revtel} we assume the following vierbein fields.

For the Bianchi I spacetime we consider the diagonal frame
\[
e^{1}=Ndt~,~e^{2}=e^{\alpha+\beta}dx~,~e^{3}=e^{\alpha-\frac{\beta}{2}%
}dy~,~e^{4}=e^{\alpha-\frac{\beta}{2}}dz.
\]

For the Kantowski-Sachs spacetime we assume%
\begin{align*}
e^{1}  &  =Ndt~,\\
e^{2}  &  =e^{a+\beta}\cos z\sin y~dx+e^{a-\frac{\beta}{2}}\left(  \cos y\cos
z~dy-\sin y\sin z~dz\right)  ~,\\
e^{3}  &  =e^{a+\beta}\sin y\sin z~dx+e^{a-\frac{\beta}{2}}\left(  \cos y\sin
z~dy-\sin y\cos z~dz\right)  ,\\
e^{4}  &  =e^{a+\beta}\cos y~dx-e^{a-\frac{\beta}{2}}\sin y~dy~,
\end{align*}
while for the Bianchi III spacetime it follows from where we calculate the
torsion scalar%
\begin{align*}
e^{1}  &  =Ndt~,\\
e^{2}  &  =i~e^{\alpha+\beta}\cos z\sinh y~dx+e^{\alpha-\frac{\beta}{2}%
}\left(  \cosh y\cos z~dy-\sinh y\sin z~dz\right)  ,\\
e^{3}  &  =i~e^{\alpha+\beta}\sinh y\sin z~dx+e^{\alpha-\frac{\beta}{2}%
}\left(  \cosh y\sin z~dy-\sinh y\cos z~dz\right)  ,\\
e^{4}  &  =-e^{\alpha+\beta}\cosh y~dx-i~e^{\alpha-\frac{\beta}{2}}\sinh y~dy,
\end{align*}

Fro the above three-different vierbein fields we derive the torsion scalar%

\begin{equation}
T=\frac{1}{N^{2}}\left(  6\dot{\alpha}^{2}-\frac{3}{2}\dot{\beta}^{2}\right)
-2Ke^{-2\alpha+\beta}. \label{ch.06}%
\end{equation}
where $K=0$\ corresponds to the Bianchi I space, $K=1$ is for Kantowski-Sachs
spacetime and $K=-1$ corresponds to the Bianchi III geometry. Moreover, $K$ is
the value of the spatially curvature in the limit the anisotropic line element
(\ref{ch.03}) becomes isotropic and the
Friedmann--Lema\^{\i}tre--Robertson--Walker \ (FLRW) geometry is recovered.
Indeed, Bianchi I spacetime is related with the spatially flat FLRW space,
while the Kantowski-Sachs and the\ Bianchi III spacetimes provide in the limit
of isotropization the closed and open FLRW spacetimes respectively.\newline

Furthermore, the boundary term $B$ is derived to be
\begin{equation}
B=\frac{6}{N^{2}}\left(  \ddot{\alpha}-\dot{\alpha}\frac{\dot{N}}{N}%
+3\dot{\alpha}^{2}\right)  , \label{ch.05}%
\end{equation}
from where we observe that~$B$ is the same for the spacetimes of our consideration.

\subsection{Minisuperspace description}

An important characteristic of the spacetimes of our consideration is that
they admit a minisuperspace description \cite{mr}. Not all cosmological
theories have a minisuperspace description. For a full-scale factor matrix and
a non-vanishing shift, the minisuperspace description admits the Bianchi class
A\ models and the Bianchi V spacetime \cite{mr}. Recall that Bianchi I,
Bianchi III and Bianchi IX belong to the family of Class A spaces.

By the minisuperspace description we mean that there exists a Lagrangian
function of the form%
\begin{equation}
S=\int\sqrt{-g}\mathcal{L}dx^{4}\rightarrow S=\int\mathcal{L}\left(
N,\mathbf{q,\dot{q}}\right)  dt,
\end{equation}
whose variation produces the gravitational field equations.

For a specific family of gravitational models, Lagrangian $\mathcal{L}\left(
N,\mathbf{q,\dot{q}}\right)  $ is of the form of a point-like Lagrangian%
\begin{equation}
\mathcal{L}\left(  N,\mathbf{q,\dot{q}}\right)  =\left[  \frac{1}{2N}%
\gamma_{\alpha\beta}\left(  \mathbf{q}\right)  \dot{q}^{\alpha}\dot{q}^{\beta
}-N\mathbf{U}\left(  \mathbf{q}\right)  \right]  ,
\end{equation}
where now $\gamma_{\alpha\beta}\left(  \mathbf{q}\right)  $ is a second-rank
tensor called the minisuperspace metric. Function $\mathbf{U}\left(
\mathbf{q}\right)  $ is called the effective potential for the given theory
which drives the dynamics.

The existence of a point-like Lagrangian for a cosmological model is an
important theory. Indeed, the point-like Lagrangian can be used to write the
field equations in Hamiltonian-formalism and perform quantization, or to apply
important techniques for the study of the field equations such as the Noether
symmetry analysis which we perform in this work.

In order to derive the minisuperspace Lagrangian we follow the procedure
applied in \cite{paper1}. We introduce the Lagrange multipliers $\lambda_{1}$
and $\lambda_{2}$ such that the gravitational Action Integral becomes%

\begin{align}
S_{f\left(  T,B\right)  }  &  =\frac{1}{16\pi G}\int d^{4}xNe^{3\alpha}\left(
f(T,B)\right)  .\label{ch.07}\\
&  +\frac{1}{16\pi G}\int d^{4}xNe^{3\alpha}\lambda_{1}\left(  T-\frac
{1}{N^{2}}\left(  6\dot{\alpha}^{2}-\frac{3}{2}\dot{\beta}^{2}\right)
+2Ke^{-2\alpha+\beta}\right) \nonumber\\
&  +\frac{1}{16\pi G}\int d^{4}xNe^{3\alpha}\lambda_{2}\left(  B-\frac
{6}{N^{2}}\left(  \ddot{\alpha}-\dot{\alpha}\frac{\dot{N}}{N}+3\dot{\alpha
}^{2}\right)  \right)  .\nonumber
\end{align}
Lagrange multiplier approach has been widely applied in cosmological studies
see for instance \cite{lan1,lan2,lan3}.

We follow \cite{paper1}, thus from the variation of (\ref{ch.07}) with respect
to the scalars $T$ and $B$ we find $\lambda_{1}=-f_{,T}$ and $\lambda
_{2}=-f_{,B}$.

Thus, by replacing in (\ref{ch.07}) we end with the point-like Lagrangian%
\begin{equation}
\mathcal{L}=\frac{1}{N}\left(  f_{,T}e^{3\alpha}\left(  6\dot{\alpha}%
^{2}-\frac{3}{2}\dot{\beta}^{2}\right)  -6e^{3\alpha}f_{,BB}\dot{\alpha}%
\dot{B}\right)  +Ne^{3\alpha}\left(  f-Tf_{,T}-BF_{,B}\right)  -2NKf_{,T}%
e^{\alpha+\beta}. \label{ch.08}%
\end{equation}

For the function $f\left(  T,B\right)  $ we consider that it is separable,
that is $f\left(  T,B\right)  =L\left(  T\right)  +F\left(  B\right)  $. Then,
the point-like Lagrangian (\ref{ch.08}) reads%
\begin{equation}
\mathcal{L}=\frac{1}{N}\left(  L_{,T}e^{3\alpha}\left(  6\dot{\alpha}%
^{2}-\frac{3}{2}\dot{\beta}^{2}\right)  -6e^{3\alpha}F_{,BB}\dot{\alpha}%
\dot{B}\right)  +Ne^{3\alpha}\left(  L-TL_{,T}\right)  +Ne^{3\alpha}\left(
F-BF_{,B}\right)  -2NKL_{,T}e^{\alpha+\beta},
\end{equation}
or equivalently%
\begin{equation}
\mathcal{L}=\frac{1}{N}\left(  L_{,T}e^{3\alpha}\left(  6\dot{\alpha}%
^{2}-\frac{3}{2}\dot{\beta}^{2}\right)  -6e^{3\alpha}\dot{\alpha}\dot{\phi
}\right)  +Ne^{3\alpha}\left(  L-TL_{,T}\right)  +Ne^{3\alpha}V\left(
\phi\right)  -2NKL_{,T}e^{\alpha+\beta}.\label{ch.10}%
\end{equation}
in which $\phi=F_{,B}$ and $V\left(  \phi\right)  =F-BF_{,B}$. Moreover, in
the following we assume $N=1$.

In \cite{paper1,paper2,paper3} we focused on the case where $L\left(
T\right)  =T$, thus we end with the Lagrangian function%
\begin{equation}
\mathcal{L}=e^{3\alpha}\left(  6\dot{\alpha}^{2}-\frac{3}{2}\dot{\beta}%
^{2}\right)  -6e^{3\alpha}\dot{\alpha}\dot{\phi}+e^{3\alpha}V\left(
\phi\right)  -2Ke^{\alpha+\beta}.\label{ch.11}%
\end{equation}

By definition in order for $F\left(  B\right)  $ to be a non-linear function,
potential $V\left(  \phi\right)  $ should be non-linear.

\section{Noether symmetry analysis}

\label{sec4}

Consider now the infinitesimal transformation with generator the vector field%
\[
X=\xi\left(  t,\alpha,\beta,\phi\right)  \partial_{t}+\eta^{1}\left(
t,\alpha,\beta,\phi\right)  \partial_{\alpha}+\eta^{2}\left(  t,\alpha
,\beta,\phi\right)  \partial_{\beta}+\eta^{3}\left(  t,\alpha,\beta
,\phi\right)  \partial_{\phi}.
\]
Hence, for the Lagrangian function (\ref{ch.11}) and the symmetry condition
(\ref{sm1}) we end with a system of partial differential equations which
constrain the \ generator $X$, the scalar field potential $V\left(
\phi\right)  $ and the curvature constant $K$. In order to solve the symmetry
conditions we follow the approach presented in \cite{p1a}. The results are
summarized in the following Proposition.

\textbf{Proposition 1:}\label{prop1} \textit{For a non-linear scalar field
potential }$V\left(  \phi\right)  $\textit{, the Lagrangian function
(\ref{ch.11}) admits for arbitrary potential the symmetry vector }%
$X^{1}=\partial_{t}$\textit{. Furthermore, when }$V\left(  \phi\right)
=V_{0}e^{-\lambda\phi}$\textit{, the Noether symmetry vector exists }%
$X^{2}=2t\partial_{t}+\frac{2}{3}\left(  \partial_{\alpha}-4\partial_{\beta
}\right)  +\frac{4}{\lambda}\partial_{\phi}$\textit{. In the special case of
Bianchi I spacetime, for arbitrary potential function }$V\left(  \phi\right)
$\textit{ there exists the additional Noether symmetry }$X^{3}=\partial
_{\beta}$\textit{. }

Therefore, with the use of Noether's second theorem, that is, from expression
(\ref{sm2}) for each symmetry vector we can construct a conservation law. The
conservation laws are as presented in the following.

\textbf{Proposition 2:}\label{prop2} \textit{The gravitational field equations
described by the point-like Lagrangian (\ref{ch.11}) admits for arbitrary
potential the conservation law }%
\begin{equation}
\mathcal{H}=e^{3\alpha}\left(  6\dot{\alpha}^{2}-\frac{3}{2}\dot{\beta}%
^{2}\right)  -6e^{3\alpha}\dot{\alpha}\dot{\phi}-e^{3\alpha}V\left(
\phi\right)  +2Ke^{\alpha+\beta},
\end{equation}
\textit{ related to the vector field }$X^{1}=\partial_{t}$\textit{. However,
}$\mathcal{H}$\textit{ is nothing else than the constraint equation, that is
}$\mathcal{H}=0$\textit{. When }$K=0$\textit{, there exist the additional
conservation law }%
\begin{equation}
I\left(  X^{3}\right)  =e^{3\alpha}\dot{\beta}.
\end{equation}
\textit{ Finally, for }$V\left(  \phi\right)  =V_{0}e^{-\lambda\phi}$\textit{,
the conservation law related to the vector field }$X^{2}$\textit{ is}%
\begin{equation}
I\left(  X^{2}\right)  =2t\mathcal{H-}\frac{2e^{3\alpha}}{\lambda}\left(
4\left(  \lambda-3\right)  \dot{\alpha}+3\lambda\dot{\beta}-2\lambda\dot{\phi
}\right)  .
\end{equation}

At this point it is important to mention that for the exponential potential
$V\left(  \phi\right)  =V_{0}e^{-\lambda\phi}$ with $\lambda=3,$ from
\cite{paper1,paper2,paper3} we know that the future attractor of the field
equations is the de Sitter Universe. We observe that the functional form of
the conservation law $I\left(  X^{2}\right)  $ depends on the parameter
$\lambda$, and when $\lambda=3$, \ with the use of the constraint equation in
the limit $I\left(  X^{2}\right)  ~$we find~$\phi\left(  t\right)  =\frac
{2}{3}\beta\left(  t\right)  +\phi_{0}$.

Noether symmetry vectors can be used to construct invariant functions. Indeed,
from $X^{2}$ we can define the invariant functions $u_{1}=\alpha-\frac{1}%
{3}\ln t$,~$u_{2}=\beta+\frac{4}{3}\ln t$ and $u_{3}=\phi-\frac{2}{\lambda}\ln
t$. $\ $However, these invariant functions are related to the exact solutions
found previously in \cite{paper2,paper3}. What it is of special interest is
the case of Bianchi I spacetime for the exponential potential. For this
cosmological model the field equations admit three conservation laws which are
independent and involution, which means that the field equations are Liouville integrable.

We proceed with the derivation of the analytic solution for the Bianchi I model.

\subsection{Bianchi I analytic solution}

For the Bianchi I spacetime and the exponential potential $V\left(
\phi\right)  =V_{0}e^{-\lambda\phi},$ let us define the lapse function
$N=e^{3\alpha}$ and the new scalar field $\psi=\phi-\frac{6}{\lambda}\alpha$.
In the new variables the Lagrangian of the field equations is
\begin{equation}
\mathcal{L}=\frac{6}{\lambda}\left(  \lambda-6\right)  \dot{\alpha}^{2}%
-\frac{3}{2}\dot{\beta}^{2}-6\dot{\alpha}\dot{\psi}+V_{0}e^{-\lambda\psi
}\text{.}%
\end{equation}
Consequently, the field equations in the new variables are
\begin{equation}
\frac{6}{\lambda}\left(  \lambda-6\right)  \dot{\alpha}^{2}-\frac{3}{2}%
\dot{\beta}^{2}-6\dot{\alpha}\dot{\psi}-V_{0}e^{-\lambda\psi}=0~,
\end{equation}%
\begin{equation}
\ddot{\alpha}-\frac{\lambda V_{0}}{6}e^{-\lambda\psi}=0~,
\end{equation}%
\begin{equation}
\ddot{\psi}-\frac{V_{0}\left(  \lambda-6\right)  }{3}e^{-\lambda\psi}=0~,
\end{equation}
and%
\begin{equation}
\ddot{\beta}=0.
\end{equation}

It is easy to see now that the additional conservation laws in the new
variables are $\dot{\beta}=const$ and $\,\frac{2\left(  \lambda-6\right)
}{\lambda}\dot{\alpha}-\dot{\psi}=const$. \ Therefore%
\begin{equation}
\beta\left(  t\right)  =\beta_{0}t+\beta_{1}~,~\alpha\left(  t\right)
=\frac{\lambda}{2\left(  \lambda-6\right)  }\psi\left(  t\right)  +\alpha
_{0}~,
\end{equation}
where now%
\begin{equation}
\psi\left(  t\right)  =-\frac{1}{\lambda}\ln\left(  2\left(  \lambda-6\right)
V_{0}\cos^{2}\left(  \frac{\sqrt{3\lambda\psi_{1}}}{6}\left(  t-t_{0}\right)
\right)  \right)  \text{.}%
\end{equation}

However, for $\lambda=6$, the conservation laws are $\dot{\beta}=0$ and
$\dot{\psi}=0$, where now the analytic solution is
\begin{equation}
\beta\left(  t\right)  =\beta_{0}t+\beta_{1}~,~\psi\left(  t\right)  =\psi
_{0}t+\psi_{1}~,
\end{equation}
and%
\begin{equation}
\alpha\left(  t\right)  =\frac{V_{0}}{216\psi_{0}^{2}}e^{-6\left(  \psi
_{0}t+\psi_{1}\right)  }+\alpha_{1}t+\alpha_{0}\text{.}%
\end{equation}

\section{Conclusions}

\label{sec5}

In this work we applied the Noether symmetry analysis to $f\left(  T,B\right)
$-theory for the field equations in anisotropic cosmological background
geometries. Specifically, we focused on the case where $f\left(  T,B\right)
=T+F\left(  B\right)  $ and we extended previous analysis on the study of the
dynamics. Noether symmetry analysis is a powerful method for the constraint of
the unknown functions and parameters of the dynamical system and the
construction of conservation laws.

In order to be able to apply the symmetry analysis the field equations should
admit a minisuperspace description. That is valid for the cosmological model
of our consideration. Indeed, the Bianchi I, the Bianchi III and the
Kantowski-Sachs geometries admit a minisuperspace. For the $f\left(
T,B\right)  =T+F\left(  B\right)  $ theory, the Lagrangian of the field
equations is point-like Lagrangian and previous studies on the analysis of the
symmetries can be applied. Thus, the Noether symmetries for the field
equations are determined by the collineations of the minisuperspace \cite{p1a}
and that gives a geometric character for the Noether symmetry analysis.

In terms of the point symmetries we found that there exists a unique function
$F\left(  B\right)  $ that is, $F\left(  B\right)  =-\frac{1}{\lambda}%
B\ln\left(  B\right)  $ where the field equations admit non-trivial Noether
symmetries. Additional, in the case of the Bianchi I model the field equations
admit a three-dimensional Lie algebra of Noether symmetries, where the
corresponding conservation laws are independent and involution, which means
that the field equations are Liouville integrable and the generic solution to
this specific model was presented in closed-form functions.

However, because for the Bianchi I spacetime the field equations are
integrable, we can investigate the existence of exact solutions and for the
Wheeler-DeWitt (WdW) equation of quantum cosmology \cite{wd1}. When a
minisuperspace description exists the WdW equation is represented as a single
differential equation \cite{ss2,wdw2,wdw5,wdw6}. Indeed, the WdW equation in
this case follows from the canonical quantization of the constraint equation.
Indeed we define the momentum $\mathbf{p}=\frac{\partial L}{\partial
\mathbf{q}}$ and we quantize as $\mathbf{p=}i\frac{\partial}{\partial
\mathbf{q}}$.

Hence, the WdW equation for our model is
\begin{equation}
\left(  \frac{1}{3}\left(  \frac{\partial^{2}}{\partial\alpha\partial\phi
}+\frac{\partial^{2}}{\partial\beta^{2}}+\frac{\partial^{2}}{\partial\phi^{2}%
}\right)  -V_{0}e^{6\alpha-\lambda\phi}\right)  \Psi\left(  \alpha,\beta
,\phi\right)  =0
\end{equation}
From the Noether symmetry vectors we can define the quantum operators $\left(
\frac{\partial}{\partial\beta}-\beta_{1}\right)  \Psi=0,~$and $\left(
\frac{\lambda}{6}\frac{\partial}{\partial\alpha}+\frac{\partial}{\partial\phi
}-\beta_{2}\right)  \Psi=0.$

Finally, with the use of the latter operators we determine the closed-form
solution%
\begin{equation}
\Psi\left(  \alpha,\beta,\phi\right)  =U\left(  \psi\right)  \exp\left(
\frac{6\beta_{2}\alpha+\beta_{1}\beta\lambda}{\lambda}\right)  ~,~\psi
=\frac{6}{\lambda}\alpha-\phi
\end{equation}
where
\begin{equation}
\left(  \lambda-6\right)  U_{,\psi\psi}+6\beta_{2}U_{.\psi}+\beta_{1}%
^{2}\lambda U-3V_{0}\lambda e^{-\lambda\psi}U=0.
\end{equation}

The latter solution is expressed in terms of the Bessel functions of the first
kind, that is,%
\begin{equation}
U\left(  \psi\right)  =e^{-\frac{3\beta_{2}}{\lambda-6}\psi}\left(
U_{1}J_{\mu}\left(  \sigma\left(  \psi\right)  \right)  +U_{2}Y_{\mu}\left(
\sigma\left(  \psi\right)  \right)  \right)  ~,
\end{equation}
where $\mu=-\frac{2\sqrt{\lambda\left(  6-\lambda\right)  +9\beta_{2}^{2}}%
}{\lambda\left\vert \lambda-6\right\vert }$, $\sigma\left(  \psi\right)
=2\sqrt{\frac{3V_{0}}{\lambda\left(  \lambda-6\right)  }}\exp\left(
-\frac{\lambda}{2}\psi\right)  $, or
\begin{equation}
U\left(  \psi\right)  =U_{1}\exp\left(  -\frac{2\beta_{1}^{2}\psi
+V_{0}e^{-6\psi}}{2\beta_{2}}\right)  ~,~\lambda=6\text{.}%
\end{equation}

We stop our discussion at this point. In a future work we plan to further
investigate the WdW equation and the quantum potentiality for the $f\left(
T,B\right)  $-theory. This work concludes a series of studies on the analysis
of the dynamics for the $f\left(  T,B\right)  $-theory in anisotropic cosmologies.

\end{document}